\documentclass[a4paper,3p]{elsarticle2}

\usepackage[T1]{fontenc}
\usepackage{graphicx}
\usepackage{amssymb}
\usepackage{amsmath,amsthm}

\newcommand{\var}{\ensuremath{\mathrm{Var}}}
\newcommand{\cov}{\ensuremath{\mathrm{Cov}}}
\newtheorem{definition}{Definition}

\begin{document}

\title{A higher order correlation unscented Kalman filter
}

\author[rvt]{Oliver~Grothe} 
\ead{grothe@statistik.uni-koeln.de}

\address[rvt]{University of Cologne, Department of Economic and Social Statistics,
Albertus-Magnus-Platz, 50923 K\"oln, Germany}

\begin{abstract}
\noindent  Many nonlinear extensions of the Kalman filter, e.g., the extended and the unscented Kalman filter, reduce the state densities to Gaussian densities. This approximation gives sufficient results in many cases. However, this filters only estimate states that are correlated with the observation. Therefore, sequential estimation of diffusion parameters, e.g., volatility, which are not correlated with the observations is not possible. While other filters
overcome this problem with simulations, we extend the measurement update of the Gaussian two-moment filters by a higher order correlation measurement update. We explicitly state formulas for a higher order unscented Kalman filter within a continuous-discrete state space. We demonstrate the filter in the context of parameter estimation of an Ornstein-Uhlenbeck process.
\end{abstract}


\date{January 2, 2012}
\maketitle


\section{Introduction}
Stochastic filters are powerful tools for simultaneous
estimation of parameters and unoberserved states from noisy data. In nonlinear filtering problems, however, in contrast to the classical linear Kalman filter setting \cite{Kalman1960}, nonlinear transformations of probability densities have to be evaluated. Since these transformations cannot be tracked analytically, approximations have to be used to construct nonlinear filters. This paper deals with a new approximation based on martingale estimation functions, which enables correlation-based nonlinear filters to estimate states which are not directly correlated with observations like, for example, stock price volatility.

A detailed survey of different nonlinear filters and their approximations is given in \cite{Tanizaki1996} (see also \cite{HermosoCarazo}). Generally, one can distinguish between simulation-based approaches and analytical approximations. While simulation-based approaches often lead to more accurate descriptions of the densities, analytical approximations are superior with respect to computation costs. The most common analytical approximation is capturing only the first two moments of the densities, i.e., assuming Gaussian densities. This strongly simplifies all filter equations and leads to computationally very fast filters. The extended Kalman Filter (see, e.g., \cite{Jazwinski1970}) and the unscented Kalman Filter (see, e.g., \cite{Julier1997}) are well-known examples for such Gaussian filters.

Assuming Gaussian densities means assuming Gaussian dependence. For this reason, these filters are only able to estimate states and latent states that are linearly correlated with the observations.
However, the cases where there is no linear correlation are common and important.
In the context of econometric modelling, is the estimation of stock price volatility a
typical example. Since price observations are not linearly correlated with the current volatility parameters, Gaussian filters are not able to sequentially estimate volatility given a sequence of price observations. Therefore other filters like simulation-based filters or particle filters have to be used (see, e.g., \cite{Pitt1999}).
However, especially in the context of high frequency financial data, computation time is a crucial point. Therefore, in many applications the use of simulation-based filters is not possible.
For these applications, fast stochastic filters are needed which are able to estimate diffusion parameters.

In this paper, we extend the filter equations of Gaussian filters by a higher order measurement update.
This measurement update relies on the theory of martingale estimating functions and uses additional information on the squared prediction error, i.e., the square of the difference between prediction of an observation value and the actual observation value. The update of normal Gaussian filters uses only the correlation of the prediction error with the states.
The correlation of the squared prediction error relies on information on higher moments of the densities, that have to be propagated in the time update steps of the filters. This can be realized in different ways.
We explicitly state an algorithm for a higher order correlation unscented Kalman filter.
We believe, that an algorithm in the sense of an higher order correlation extended Kalman filter is also possible to implement, but relies on quite complicated moment equations which are not necessary for the unscented filter. For the unscented filter we use an extended version of the unscented transformation to capture the information on higher moments.
We demonstrate the applicability of the algorithm in a simulation study estimating the parameters of an Ornstein-Uhlenbeck process.

The paper is organized as follows. In section \ref{sec.nonl_state_est}, nonlinear state estimation in the framework of continuous-discrete state space models is discussed deriving the
general filter equations for Gaussian filters and introducing a higher order correlation update.
In section \ref{sec.ukf}, higher order filter equations for the unscented Kalman filter are derived. Section \ref{sec.simustud} contains a simulation study analyzing bias and variance of the filter estimates in the context of an Ornstein-Uhlenbeck model. Section \ref{sec.concl} concludes.

\section{Continuous discrete higher order correlation filters}
\label{sec.nonl_state_est}
\subsection{State space models and parameter estimation}
We concentrate on continuous discrete state space models (see \cite{Jazwinski1970}) as they are very useful in
systems, in which the underlying models are continuous in time and in which
only discrete observations are available. Since these models are more general than discrete state space models, the results in this paper also hold for the discrete case.

The continuous discrete state space consists of a
continuous state equation for the state $y(t)$ (equation (\ref{glg.kontidisk}))  and discrete
measurements $z_i$ at times $t_i, i=1\dots n,$ which do not have to be evenly spaced in time (equation (\ref{glg.kontidisk2})):
\begin{eqnarray}
 \label{glg.kontidisk}
dy(t)&=&f(y(t),t,\Psi)dt+g(y(t),t,\Psi)dW(t),
\\ \label{glg.kontidisk2} z_i&=&h(y(t_i),t_i)+{\epsilon}_i.
\end{eqnarray}
The first equation is a $p$-di\-men\-sio\-nal It\^{o} differential
equation with an $r$-di\-men\-sio\-nal Wiener process $W(t)$. The
drift coefficient
$f:\mathbb{R}^p\times\mathbb{R}\times\mathbb{R}^u\longrightarrow\mathbb{R}^p$
and the diffusion coefficient
$g:\mathbb{R}^p\times\mathbb{R}\times\mathbb{R}^u\longrightarrow\mathbb{R}^p\times\mathbb{R}^r$
are functions of the state, the time and a $u$-dimensional
parameter vector $\Psi$. The measurement equation projects the state
vector $y(t)$ onto the time discrete $k$-dimensional measurements
$z_i$. The measurement may be noisy with the $k$-dimensional
discrete white noise process
$\epsilon_i\sim{N}(0,R(t_i,\Psi))$, $\epsilon_i,$ identically distributed and
independent of $W(t)$.

The state space notation has an inherent way of parameter estimation by considering the parameter vector $\Psi$ as a latent state
variable. The state vector is augmented by the parameter vector
($y \rightarrow {y} = {y \brack \Psi}$) with trivial
dynamics ($d\Psi=0$). This leads to the extended state space
model:
\begin{eqnarray*} dy(t)&=&f(y(t),t,\Psi)dt+g(y(t),t,\Psi)dW(t)
\\ \nonumber d\Psi&=&0
\\ \nonumber z_i&=&h(y(t_i),t_i)+\tilde{\epsilon}_i.
\end{eqnarray*}
Filtering this state space model results in a
sequential estimator of the unknown parameters:
$\hat{\Psi}=E[\Psi(t)|Z^t]$. Note, that this approach of parameter estimation leads to nonlinear problems, even if the original state equations are linear. Therefore, linear filters
like the Kalman filter
are inapplicable for this approach.
Standard nonlinear
 filters as the extended Kalman filter or the unscented Kalman filter are applicable
but do not estimate parameters of the diffusion coefficient
$g(y(t),t,\Psi),$ e.g., volatility.

\subsection{Time and measurement update for Gaussian filters}
\label{sec.filters}

Continuous discrete stochastic filters estimate the probability
density of the state vector from noisy observations. The algorithms consist of two steps. There is no measurement information for
$t\in]t_i,t_{i+1}[.$ All information is based on time $t_i$ and the density
$p(y,t|Z^i)$ is obtained by propagating $p(y,t_i|Z^i)$ according to the dynamic given by the state equation (\ref{glg.kontidisk}). This results in integrating the Fokker-Planck equation (see, e.g.,
\cite{Jazwinski1970}). This step is the \emph{time update}.
For $t=t_{i+1},$ with current measurement
information available, the a-posteriori (after measurement) state density
$p(y,t_{i+1}|Z^{i+1})$ is estimated via the Bayes' theorem using the a-priori density $p(y|Z^i)$ of the time update step
and the measurement information $p(z_{i+1})$ and likelihood $p(z_{i+1}|y).$
This step is the \emph{measurement update}. Time update and measurement update are recursively repeated. Note that both steps can be solved explicitly only for linear systems and Gaussian densities and some special cases. Therefore, various numerical and simulation-based methods exist to approximately solve both steps (see for example \cite{Singer2003} and the reviewing introduction therein).

A common simple and effective approximation is assuming Gaussian densities although the transformations are nonlinear. This simplifies the equations and is done, e.g., by the extended Kalman filter (EKF, \cite{Jazwinski1970}) and the unscented Kalman filter (UKF, \cite{Julier2004}). These Gaussian filters only estimate mean $\mu$ and variance $\Sigma$ of the state densities and are exact for linear problems, i.e., linear transformations of the densities. The Fokker-Planck equation of the time update step then reduces to the moment equations
\begin{eqnarray} \label{glg.timeupd1} \dot{\mu}(t)&=&E[f(y(t),t)|Z^i]  \\
\label{glg.timeupd2}
\dot{\Sigma}(t)&=&\mathrm{Cov}[f,y|Z^i]+\mathrm{Cov}[y,f|Z^i]+E[\Omega|Z^i],
\end{eqnarray}
with $\Omega=gg'$. A dot over the variables denotes differentiation with respect to time. Since these equations depend on the conditional
densities $p(y,t|Z^i),$ they have to be solved approximately.
The extended Kalman filter uses linear Taylor expansions to approximate the functions $f$
and $g,$ while the unscented Kalman filter approximates the densities of $f(y)$ and $g(y).$
The measurement update is simplified using the theorem of normal
correlation for jointly distributed Gaussian random variables.\footnote{Let $X$ and $Y$ be normally distributed random vectors with expectations $\mu_X, \mu_Y$ and covariances $\Sigma_{XX}, \Sigma_{YY}, \Sigma_{XY}, $ respectively. Then $E[X|Y]=\mu_X+\Sigma_{XY}\Sigma_{YY}^{-}(Y-\mu_Y)$ and $\var[X|Y]=\Sigma_{XX}-\Sigma_{XY}\Sigma_{YY}^{-}\Sigma_{YX}, $ where $\Sigma_{YY}^{-}$ is the pseudoinverse of $\Sigma_{YY}.$} Using subscripts $_i$ for \emph{at
time $t_i$} and $_{i+1|i}$ for \emph{at time $t_{i+1}$ based on
information of $t_i$} and taking the measurement equation (\ref{glg.kontidisk2}) into
account with $h_i:=h(y(t_i)),$ the measurement update becomes
\begin{eqnarray}
\label{glg.measupd1}\mu_{i+1|i+1}&=&\mu_{i+1|i}+K_{i+1}(z_{i+1}-E[h_{i+1}|Z^i]), \\
\label{glg.measupd2}
\Sigma_{i+1|i+1}&=&\Sigma_{i+1|i}-K_{i+1}\cov[h_{i+1},y_{i+1}|Z^i],
\end{eqnarray}
where \begin{align}K_{i+1}=\cov[y_{i+1},h_{i+1}|Z^i] (\var[h_{i+1}|Z^i]+R_{i+1})^{-} \label{glg.kalmangain}\end{align} is a regression matrix, called Kalman gain, and $(\cdot)^{-}$ denotes the pseudoinverse.
Expectations and covariances have to be
approximated. The extended Kalman Filter uses Taylor expansions of the respective functions around
$\mu_{i+1|i}$ and the unscented Kalman filter uses the unscented transformation to approximate the transformed densities (see section \ref{sec.ukf}).

Due to the role of the covariance matrices $\cov[y_{i+1},h_{i+1}|Z^i]$ in equations (\ref{glg.measupd1}) and (\ref{glg.measupd2}), only the states which are correlated with the measurement are updated.
States (or parameters) that are connected to the measurement through the diffusion coefficient $g$, e.g., volatility parameters, are generally not correlated with the measurement, i.e., they are not updated/estimated by the filter. In the next section, we extend both steps with a higher order correlation component to overcome this problem.

\subsection{Higher order correlation measurement update}
States which are not linearly correlated with the measurement are not
updated via the theorem of normal correlation and cannot be
estimated by Gaussian filters. In these cases, the corresponding entries in the
covariance matrix $\cov[y_{i+1},h_{i+1}|Z^i]$ are zero and
the prediction error $\nu_{i+1}=(z_{i+1}-E[h_{i+1}|Z^i])$ in equation (\ref{glg.measupd1}) does not provide
information for a linear update of these states.
However, in most cases the squared prediction error
$\nu_{i+1}^2=(z_{i+1}-h_{i+1|i})^2$ provides additional information on these states.
Therefore, a measurement update is constructed in this section which also consists of a nonlinear part based on the squared prediction error.

The measurement update in equation (\ref{glg.measupd1}) is closely related to the theory of martingale estimating functions and quasi-likelihood models for stochastic models. If  $\theta$ is a real valued random variable on a probability space $(\Omega=\{\omega\}, \mathcal{F}, \mathcal{P}),$ an estimating function for $\theta$ is a function $G=g(\omega, \theta)$ on $\Omega\times\mathbb{R}.$ It is unbiased if $E_{\mathcal{P}}\left [g(\omega, \theta(\omega))\right] =0,$ where $E_{\mathcal{P}}$ is the expectation over $\mathcal{P}$ (see, e.g., \cite{Godambe1991} for a survey).

For example, let $X_i,$ $i=1\dots n$ be $n$ independent random variables with finite mean $m_i(\theta)$ and variance $\upsilon_i(\theta)$. Then $E[X_i-m_i(\theta)]=0$ since the sequence $X_i-m_i(\theta)$ is a martingale difference.
Since $E\left[\left(X_i-m_i(\theta)\right)^2\right]=\upsilon_i(\theta),$
$$E\left[\left(X_i-m_i(\theta)\right)^2-\upsilon_i(\theta)\right]=0$$ holds in the same way, which yields the estimating functions for $\theta:$
\begin{align} 0&\stackrel{!}{=}\sum_{i=1}^n\omega_{1,i}\left(X_i-m_i(\theta)\right), \label{glg.martin1} \\  0&\stackrel{!}{=}\sum_{i=1}^n\omega_{2,i}\left[\left(X_i-m_i(\theta)\right)^2-\upsilon_i(\theta)\right], \label{glg.martin2} \end{align}
where $\omega_{1/2,i}$ are weights. Together with the observation equation (\ref{glg.kontidisk2}), the linear estimating function (\ref{glg.martin1}) directly leads to the linear filter equations (\ref{glg.measupd1}) and (\ref{glg.measupd2}). See, e.g., \cite{NaikNimbalkar1995} or \cite{Thompson1999} for a derivation.

Similar to the problem considered in this paper, i.e., the failing of the linear measurement update, there are problems, where linear estimating functions of the form (\ref{glg.martin1}) fail to estimate the value of $\theta.$ An early reference with examples for this effect is \cite{Crowder1987}.
A solution is to combine equations (\ref{glg.martin1}) and (\ref{glg.martin2}) to one single estimating function
\begin{align} 0\stackrel{!}{=}\sum_{i=1}^n\left\{\omega_{1,i}^*\left(X_i-\mu_i(\theta)\right) +\omega_{2,i}\left[\left(X_i-\mu_i(\theta)\right)^2-\sigma^2_i(\theta)\right]\right\}, \label{glg.martin3}  \end{align}
to overcome this problem. Much work has been done to find the new optimal weights $\omega_{1,i}^*$ and $\omega_{2,i}^*,$ see, e.g., \cite{Crowder1987,Crowder1986} or \cite{Godambe1989137} for the restriction of orthogonality of equations (\ref{glg.martin1}) and (\ref{glg.martin2}) and \cite{Heyde1987} for the general case.

Wefelmeyer \cite{Wefelmeyer1996} showed that this estimator (\ref{glg.martin3}) may be replaced by the one-step estimator
\begin{align*} \nonumber \widehat{\theta}_n={\theta}_n+I_{n}^{-1}n^{-1}\sum_{i=1}^{n}C_{i-1}^{-1}\Bigg(A_{\theta_{n,i-1}}\left(X_i-m_{\theta_n}(X_{i-1})\right) \\ +B_{\theta_{n,i-1}}\left(\left(X_i-m_{\theta_n}(X_{i-1})\right)^2-\upsilon_{\theta_{n,i-1}}(X_i)\right)\Bigg), 
\end{align*}
where $\theta_n$ is an initial estimator for $\theta.$ The weights $A,$ $B,$ $C,$ and $I$ are
\begin{align*}
A_{\theta,i}&=m'_{\theta}(X_i)(\mu_{4,i}-\mu_{2,i}^2)-\upsilon'_{\theta}(x_i)\mu_{3,i}, \\
B_{\theta,i}&=\upsilon'(X_i)\mu_{2,i}-m'_{\theta}(X_i)\mu_{3,i}, \\
C_i&=(\mu_{4,i}-\mu_{2,i}^2)\mu_{2,i}-\mu_{3,i},\\
I_n&=n^{-1}\sum_{i=1}^nC^{-1}_{i-1}\left(A_{\theta,i-1}m'_{\theta_n}(X_{i-1})+B_{\theta,i}\upsilon'_{\theta_n}(X_{i-1})\right),
 \end{align*}
see, e.g., \cite{Heyde1987}.
 The $\mu_1, \mu_2, \mu_3$ are centered moments, $m_{\theta_n}(X_{i-1})$ is a predictor of $X_i$ based on $\theta_n$ and $X_{i-1},$ and $\upsilon_{\theta_n}(X_{i-1})$ is a predictor of the variance of $X_i.$ The primes $^\prime$ denote derivatives with respect to $\theta.$
For $n=1,$ i.e., one observation time and assuming $\mu_3=0$ for simplicity, i.e., orthogonality of equations (\ref{glg.martin1}) and (\ref{glg.martin2}),
the weights reduce to:
\begin{align}
I^{-1}C^{-1}A&=\left(m'(\mu_4-\mu_2)^2m'+\upsilon'\mu_2\upsilon'\right)^{-1}m'(\mu_4-\mu_2^2)\approx
m'^{-1}=\frac{\partial \theta}{\partial m}, \label{glg.weights1}
\\
I^{-1}C^{-1}B&=\left(m'(\mu_4-\mu_2)^2m'+\upsilon'\mu_2\upsilon'\right)^{-1}\upsilon'\mu_2\approx \upsilon'^{-1}=\frac{\partial \theta}{\partial \upsilon},
\label{glg.weights2}\end{align}
where we make the further assumption that the terms $\left(m'(\mu_4-\mu_2)^2\right)^{-1}\upsilon'\mu_2\upsilon'$ respectively  $\left(\upsilon'\mu_2\right)m'(\mu_4-\mu_2)^2m'$ are negligible. In both assumptions we assume that either the linear information or the quadratic information on $\theta$ dominates and not both terms provide the same level of information.

With the estimator $\mu_{i+1|i}$ for $\mu_{i+1|i+1}$ (notation as is section \ref{sec.filters}) this yields a new measurement update equation for the mean:
\begin{align} \label{glg.measupd_neu_1}
 \mu_{i+1|i+1}&= \mu_{i+1|i}+K_{i+1}^{(1)} \times\nu_{i+1}+K_{i+1}^{(2)} \times(\nu_{i+1}^2-E[\nu_{i+1}^2|Z^i]),
\end{align}
where $\nu_{i+1} =(z_{i+1}-E[h_{i+1}|Z^i])$ is the observation error and $K_1$ and $K_2$ are linear regressor matrices, with
\begin{align*} K_{i+1}^{(1)}&=\cov[y_{i+1},h_{i+1}|Z^i]
\left(\var[h_{i+1}|Z^i]+R_{i+1}\right)^{-}, 
\\ K_{i+1}^{(2)}&=\cov[y_{i+1},\nu^2_{i+1}|Z^i]\left(\var\left[\nu^2_{i+1}|Z^i\right]\right)^{-}, 
\end{align*}
where the derivatives in equations (\ref{glg.weights1}) and (\ref{glg.weights2}) are replaced by the respective regressors, i.e., covariances divided by variances. $K_1$ and $K_2$ may be referred to as first and second order Kalman Gain, respectively, where $K_1$ coincides with the classical Kalman Gain given in equation (\ref{glg.kalmangain}).

The equation for the variance after measurement (\ref{glg.measupd2}) is extended analogously leading to
\begin{align} \Sigma_{i+1|i+1}&= \Sigma_{i+1|i}-K_{i+1}^{(1)}\cov[h_{i+1},y_{i+1}|Z^i]-K_{i+1}^{(2)}\cov[\nu^2_{i+1},y_{i+1}|Z^i]. \label{glg.measupd_neu_2}\end{align}
Note that this approach for the variance is fairly rough. It follows from the assumption of a joint normal distribution of squared prediction errors and states. This implies independence of the results of the measurement and its information content, i.e., only the fact that there is new information and not the value of the measurement contributes to the variance update. In the case of diffusion parameters, however, small prediction errors are likely for all values of the diffusion parameters, whereas large errors clearly indicate large parameters. Therefore, large prediction errors contain more information on the diffusion coefficient than small errors. In a more accurate approximation, the value of $\nu$ would therefore be part of the variance update (\ref{glg.measupd_neu_2}).

The higher order measurement update equations (\ref{glg.measupd_neu_1}) and (\ref{glg.measupd_neu_2}) contain higher moments of the state vector $y,$ i.e., a forth centered moment and the covariance matrix $\cov[y_{i+1},\nu^2_{i+1}|Z^i]$ which contains mixed third order moments of the state vector $y.$ Thus, the time update steps of the filter algorithms have to be extended by a propagation of the respective information. In the next section, we will discuss how the update can be applied within the unscented Kalman filter framework.

\section{Higher order correlation unscented Kalman filter}
\label{sec.ukf}

The unscented Kalman filter as introduced by Julier and Uhlmann \cite{Julier1997} is based on the unscented transformation which is briefly reviewed in subsection \ref{sec.ut}. The unscented transformation captures the first two moments of densities undergoing nonlinear transformations. Since information on higher moments is needed, the unscented transformation has to be extended to capture these moments which is done in subsection \ref{sec.ut2}. In subsection \ref{sec.sshukf}, a higher order unscented Kalman filter is presented.

\subsection{Unscented transformation}
\label{sec.ut}

\label{sec.unsctrafo} The unscented transformation is a method for calculating the
transformation of the density of a random variable which undergoes
a nonlinear transformation (see \cite{Julier1997}).
In the original framework of \cite{Julier1997},
the density $p(y)$ of the random variable $y \in
\mathbb{R}^{{p}}$ is approximated by the sum
\begin{eqnarray*}
p_{UT}(y)=\sum\limits_{j=-{p}}^{{p}}\omega^{(j)}\delta(y-y^{(j)}),
\end{eqnarray*}
with the Dirac delta function $\delta,$ the $n=2{p}+1$
\emph{sigma points} $y^{(j)},$ and the weights $\omega^{(j)}$.  Furthermore the sigma points and weights are chosen such that the
first two moments of the density of $y$ are replicated:
\begin{eqnarray*}
E[y]&=&\int y p(y) dy \stackrel{\textnormal{!}}= \int y p_{UT}(y) dy = \sum_{j=1}^{n}\omega^{(j)}y^{(j)}, \\
\var[y]&\stackrel{\textnormal{!}}=&\sum_{j=1}^{n}\omega^{(j)}(y^{(j)}-E[y])(y^{(j)}-E[y])^T.
\end{eqnarray*}
In contrast to Monte Carlo approaches, this choice is
deterministic. Julier and Uhlmann \cite{Julier1997} establish the following choice:
\begin{align*}
&y^{(i)}=E[y]+\left(\sqrt{({p}+{\kappa})\var[y]}\right)_i,&&\omega^{(i)}=\frac{1}{2({p}+{\kappa})},&i=1\cdots{p},\\
&y^{(i+{p})}=E[y]-\left(\sqrt{({p}+{\kappa})\var[y]}\right)_i,&&\omega^{(i+{p})}=\frac{1}{2({p}+{\kappa})},&i=1\cdots{p},
\\ &y^{(2{p}+1)}=E[y],&&\omega^{(2{p}+1)}=\frac{{\kappa}}{({p}+{\kappa})},& \end{align*}
 where $(\sqrt{\,...\,})_i$ is the $i$-th row or column of the matrix root. The real parameter $\kappa$ gives an
 extra degree of freedom which is discussed in detail in \cite{Julier1997}. For Gaussian
 densities of $y,$ they recommend ${p}+\kappa=3.$ In many cases, however, $\kappa=0$ works well (see, e.g., \cite{Julier2004}) which reduces the number of sigma points by one.
 After undergoing a nonlinear transformation
$y\rightarrow\tilde{f}(y),$ the expectation, the covariance of
$\tilde{f}(y)$ and the cross covariance of $\tilde{f}(y)$ and $y$ can
be computed by:
\begin{eqnarray} \nonumber
E[\tilde{f}(y)]&=&\sum_{j=1}^{n}\omega^{(j)}\tilde{f}(y^{(j)}), \\ \nonumber
\var[\tilde{f}(y)]&=&\sum_{j=1}^{n}\omega^{(j)}\left(\tilde{f}(y^{(j)})-E[\tilde{f}(y)]\right)\left(\tilde{f}(y^{(j)})-E[\tilde{f}(y)]\right)',
\\ \label{glg.utcovberechnen}
\cov[\tilde{f}(y),y]&=&\sum_{j=1}^{n}\omega^{(j)}\left(\tilde{f}(y^{(j)})-E[\tilde{f}(y)]\right)\left(y^{(j)}-E[y]\right)',
\end{eqnarray}
where $(...)(...)'$ denotes the outer product.

\subsection{Higher order unscented transformation }
\label{sec.ut2}

To capture the state densities with higher accuracy, the set of sigma points of the unscented transformation may be extended (see, e.g., \cite{Julier2004}). In the following, we present an approach based on numerical integration rules for exact monomials given by \cite{McNamee1967} and \cite{Lerner2002} to derive an extended set of sigma points and weights.
\begin{definition}[Generator, \cite{Lerner2002}] A point $\mathbf{u}=(u_1,...,u_m,0,...,0)\in \mathbb{R}^p$ where $0<u_i\leq u_{i+1}$ is called generator and denoted as $[\pm \mathbf{u}]$ or $[\pm u_1,...,\pm u_m]$. It represents the set of points that can be obtained from $\mathbf{u}$ by permutations and changing the sign of some coordinates.  \end{definition}
For example, the generator $[0]$ contains only the point $0\in\mathbb{R}^p,$ while the generator $[\pm^1,\pm1]$ in $\mathbb{R}^3$ is the set of points \begin{align*} \{\,\,&(1,1,0),(-1,-1,0),1,-1,0),(-1,1,0)(1,0,1),(-1,0,-1),\\ &(1,0,-1),(-1,0,1),(0,1,1),(0,-1,-1),(0,1,-1),(0,-1,1)\,\, \}. \end{align*}
In the sense of \cite{Lerner2002}, we write $f[\pm u_1,\dots,\pm u_r]$ for the sum $\sum_{x\in[\pm u_1,\dots,\pm u_r]}f(x).$

In generator notation, the unscented transformation of Julies and Uhlmann \cite{Julier1997} with $n=2p+1$ sigma points and initial $N(0,I)$ Gaussian densities corresponds to an integration of the form $$\text{E}[f(x)]=\int_{-\infty}^{\infty}N(x;0,I)f(x)dx\approx\omega_0f[0]+\omega_1f[\pm u],$$ with the weights $\omega_k$ and the generators $[0]$ and $[\pm \sqrt{3}]$.

This integration rule is of precision $3$, i.e., it is exact for all monomials of degree $3$ or less without any odd power. To derive higher accuracy in the filtering algorithm, we use a rule of the form: \begin{align*} \int_{-\infty}^{\infty}N(x;0,I)f(x)dx\approx\omega_0f[0]+\omega_1f[\pm u]+\omega_2f[\pm u,\pm u]. \end{align*} This rule leads to $n=2p^2+1$ sigma points as we get $1$ point from the generator $[0]$, $2p$ points from $[\pm u]$ and $2p(p-1)$ points from $[\pm u,\pm u]$. For $u=\sqrt{3}$, $\omega_0=1+\frac{p^2-7p}{18}$, $\omega_1=\frac{4-p}{18}$ and $\omega_2=\frac{1}{36}$ the rule is exact for monomials of degree 5 or less without any odd power (see, e.g., \cite{Lerner2002} and \cite{McNamee1967}).

These sigma points capture integration with respect to standard normal Gaussian densities $N(0,I)$. General Gaussian densities $N(\mathbf{\mu},\Sigma)$ are captured by multiplying each sigma point with a square root of the covariance matrix $\Sigma$ and adding the mean $\mathbf{\mu}$.
Finally we get the following $2p^2+1$ sigma points and weights:
\begin{align*}
&y^{(i)}=E[y] ,&&\omega^{(i)}=1+\frac{p^2-7p}{18},&i~\text{corresp.}~[0],\\
&y^{(i)}=E[y]+\sqrt{\Sigma}\cdot[\pm\sqrt{3}]^{(i)},&&\omega^{(i)}=\frac{4-p}{18},&i~\text{corresp.}~[\pm \sqrt{3}],
\\ &y^{(i)}=E[y]+\sqrt{\Sigma}\cdot[\pm\sqrt{3},\pm\sqrt{3}]^{(i)},&&\omega^{(i)}=\frac{1}{36},&i~\text{corresp.}~[\pm \sqrt{3},\pm \sqrt{3}]. \end{align*}

The computation of moments and higher moments of transformed densities is performed analogously to equations (\ref{glg.utcovberechnen}).

\subsection{Filter algorithm}
\label{sec.sshukf}

The higher order unscented transformation can be used to evaluate the expectation,
variance and covariance terms of the filter equations leading to a higher order unscented filter.
The time
update is done by numerically solving the moment equations
(\ref{glg.timeupd1}) and (\ref{glg.timeupd2})
 by Euler integration of the sigma point sets.
With regard to the
time continuous nature of the state equations of the system, the
Euler scheme may use a finer discretization interval $\delta t \rightarrow 0$ than
the measurement intervals $t_{i+1}-t_i$ dividing them into
$L_i=(t_{i+1}-t_i)/\delta t$ parts. For notational convenience, we use the function $\tilde{f}$ in the algorithm as an abbreviation of the Euler steps.
The noise term is discretized on the Euler grid, i.e., $dW \rightarrow \Delta
W \sim \mathcal{N}(0,Q)$ with $Q\in \mathbb{R}^{r\times r}.$
The  $p$-dimensional state vector is then augmented by the $r$-dimensional noise terms $\Delta
W $ ($\tilde{p}=p+r$) to
\begin{align*}  \tilde{y}={y \brack \Delta W}. \end{align*} This leads to
$n=2(p+r)^2+1=2\tilde{p}^2+1$ $\sigma$-points based on
$$E[\tilde{y}]={E[y]\brack
0}\,\,\,\,\text{und}\,\,\,\,\Sigma_{\tilde{y}}={\begin{pmatrix}
  \Sigma_{y} & \cov(y,\Delta W):=0 \\
  \cov(\Delta W,y):=0 & Q
  \end{pmatrix}}.$$ The noise terms are uncorrelated with the states ($\cov[y,\Delta W]:=0$).

After the time update, expectations, variances and
covariances are evaluated according to equations (\ref{glg.utcovberechnen}). The measurement update is done by straight forward implementation of the equations (\ref{glg.measupd_neu_1}) and
(\ref{glg.measupd_neu_2}). The following algorithm summarizes the resulting filter:

\begin{flushleft} \textbf{Algorithm 1} (Higher order correlation unscented Kalman filter (HUKF)).   \emph{} \hspace*{0.5cm}\\
\vspace{0.2cm} \text{Initialization:
}$t=t_0$ \begin{eqnarray*} \mu_{0|0}&=&\mu+\cov[y_0,h_0]\times \\
&\times&
(\var[h_0]+R_0)^{-}\cov[h_0,y_0] \\
\Sigma_{0|0}&=&\Sigma-\cov[y_0|h_0]\times
\\&\times&(\var[h_0]+R_0)^{-}\cov[h_0,y_0] \\
L_0&=&\phi(z_0;E[h_0],\var[h_0]+R_0)\\
\text{Sigma points} & : & y^{(j)}=y^{(j)}(\mu,\Sigma),\,\,\,j=1\dots n;
\mu=E[y_0],\Sigma=\var[y_0]. \end{eqnarray*} \text{Recursion: }
$i=0,...,T-1$  \\ \vspace{0.2cm} \emph{}\hspace{1cm}\text{Time
update: }$t\in[t_i,t_{i+1}]$
\begin{eqnarray*}
y_{i+1|i}&=&\sum_{j=1}^{n}\omega^{(j)}\tilde{f}(y^{(j)},t_i,\Delta
t_i,\Delta W_{i+1}^{(j)}) \\
\Sigma_{i+1|i}&=&\sum_{j=1}^{n}\omega^{(j)}\left(\tilde{f}(y^{(j)},t_i,\Delta
t_i,\Delta W_{i+1}^{(j)})-y_{i+1|i}\right)\times \\
&&\times\left(\tilde{f}(y^{(j)},t_i,\Delta
t_i,\Delta W_{i+1}^{(j)})-y_{i+1|i}\right)'\\
E[h_{i+1}|Z^i]&=&\sum_{j=1}^{n}\omega^{(j)}h(\tilde{f}(y^{(j)},t_i,\Delta
t_i,\Delta W_{i+1}^{(j)}))=:h_{i+1|i}\\
\var[h_{i+1}|Z^i]&=&\sum_{j=1}^{n}\omega^{(j)}\left(h(\tilde{f}(y^{(j)},t_i,\Delta
t_i,\Delta W_{i+1}^{(j)}))-h_{i+1|i}\right)^2\\
E[\nu_{i+1}^2|Z^i]&=&\var[h_{i+1}|Z^i]+R_{i+1}=:\nu^2_{i+1|i} \\
\var[\nu^2_{i+1}|Z^i]&=&\sum_{j=1}^{n}\omega^{(j)}\Big[\left(h(\tilde{f}(y^{(j)},t_i,\Delta
t_i,\Delta W_{i+1}^{(j)}))-h_{i+1|i}\right)^2-\nu^2_{i+1|i}+R_{i+1}\Big]^2\\
\cov[y_{i+1},h_{i+1}|Z^i] &=& \sum_{j=1}^{n}\omega^{(j)}\left(\tilde{f}(y^{(j)},t_i,\Delta
t_i,\Delta W_{i+1}^{(j)})-y_{i+1|i}\right)\times \\
&&\times \left(h(\tilde{f}(y^{(j)},t_i,\Delta
t_i,\Delta W_{i+1}^{(j)}))-h_{i+1|i}\right) \\
\cov[y_{i+1},\nu^2_{i+1}|Z^i]&=&\sum_{j=1}^{n}\omega^{(j)}\left(\tilde{f}(y^{(j)},t_i,\Delta
t_i,\Delta W_{i+1}^{(j)})-y_{i+1|i}\right)\times \\
&&\times\left[\left(h(\tilde{f}(y^{(j)},t_i,\Delta t_i,\Delta
W_{i+1}^{(j)}))-h_{i+1|i}\right)^{2}-\nu^2_{i+1|i}\right]'\\
\text{Sigma points} & : & y{(j)}=y^{(j)}(\mu_{i|i},\Sigma_{i|i}),\,\,\,j=1\dots n
\end{eqnarray*}
\emph{}\hspace{1cm}\text{Measurement Update}
\begin{eqnarray*}
K_{i+1}^{(1)}&=&\cov[y_{i+1},h_{i+1}|Z^i] (\var[h_{i+1}|Z^i]+R_{i+1})^{-}\\
K_{i+1}^{(2)}&=&\cov[y_{i+1},\nu^2_{i+1}|Z^i] (\var[\nu^2_{i+1}|Z^i])^{-}\\
\nu_{i+1}&=&(z_{i+1}-h_{i+1|i}) \\
\mu_{i+1|i+1}&=&\mu_{i+1|i}+K_{i+1}^{(1)}\nu+K_{i+1}^{(2)}(\nu_{i+1}^2-E[\nu_{i+1}^2|Z^i]) \\
\Sigma_{i+1|i+1}&=&\Sigma_{i+1|i}-K_{i+1}^{(1)}\cov[h_{i+1},y_{i+1}|Z^i]-K_{i+1}^{(2)}\cov[\nu^2_{i+1}, y_{i+1}|Z^i]\\
\end{eqnarray*}
\emph{}\hspace{1cm}\text{with the notations}
\begin{eqnarray*}
\tilde{f}(y_{i|i},t_i,\Delta t_i,\Delta
W_i)&=&y_{i|i}+f(y_{i|i},t_i)\Delta t_i + g(y_{i|i},t_i)\Delta
W_i\sqrt{\Delta t_i} \\ & &\text{or on finer grid} \\ \{\omega^{(j)},y_{i|i}^{(j)},\Delta W_i^{(j)}\}&& \text{see text} \\
(\cdots)^2&& \text{for vectors is element wise}
\end{eqnarray*}
\end{flushleft}

$h$ is a short form for the measurement equation $h(y)$. The subscript $_{i|i}$ denotes \emph{at time $t_i$ based on information at time $t_i$}, whereas the subscript $_{i+1|i}$ denotes
\emph{at time $t_{i+1}$ based on information at time $t_{i}$}.

\section{Simulation study}
\label{sec.simustud}

In this section, we compare the presented higher order unscented Kalman filter to the unscented Kalman filter in a simulation study. The aim of the study is to show that the higher order terms in the filter equation lead to adequate estimates of states, which are correlated as well as uncorrelated to the observations.

We apply the filter algorithms to the estimation of parameters of an Ornstein-Uhlenbeck process. Ornstein-Uhlenbeck processes are regularly used to model commodity prices (see, e.g.,
\cite{Schwartz1997}), electricity prices (see, e.g.,
\cite{Lucia2002}) or interest rates
(see, e.g., \cite{Cox1985}). A
continuous discrete state space representation is:
\begin{eqnarray*}
dy(t)&=&\Psi_1\left[\Psi_2-y(t)\right]dt+\Psi_3 dW(t)
\\ \nonumber z_i&=&y(t_i)
\end{eqnarray*}
with the parameter set $\Psi=[\Psi_1,\Psi_2,\Psi_3]$.
As before, the model consists of a time continuous state equation and a discrete
measurement equation, where the observations $z_i$ are the prices or interest rates at
certain points $t_i$ in time and $y$ corresponds to a latent underlying process.
The parameter $\Psi_1$ is the mean reversion rate, $\Psi_2$ is the long run mean of $y,$ and $\Psi_3$ corresponds to the volatility of the process.
 For
simplicity, we suppress all units. Observation intervals $t_{i+1}-t_i$ as well as the intervals of the euler grids are set to $1$. To sequentially estimate the parameter vector as discussed in section \ref{sec.nonl_state_est}, we augment the state vector $y(t)$ according to
$\tilde{y}(t)=[y(t)',\Psi]'$ and get an extended state space model
\begin{eqnarray}  \label{glg.zustmodellsimu2}
d\tilde{y}_1=dy\phantom{_3}&=&\tilde{y}_2\left[\tilde{y}_3-\tilde{y}_1(t)\right]dt+\tilde{y}_4 dW(t) \\
\nonumber d\tilde{y}_2=d\Psi_1&=&0 \\ \nonumber
d\tilde{y}_3=d\Psi_2&=&0 \\ \nonumber d\tilde{y}_4=d\Psi_3&=&0
\\ \nonumber z_i&=&\tilde{y}_1(t_i).
\end{eqnarray}

\begin{figure*} [p!] \begin{center}
\makebox{ \begin{minipage}{0.65\linewidth} \begin{center}
\resizebox{\linewidth}{!}{\includegraphics*{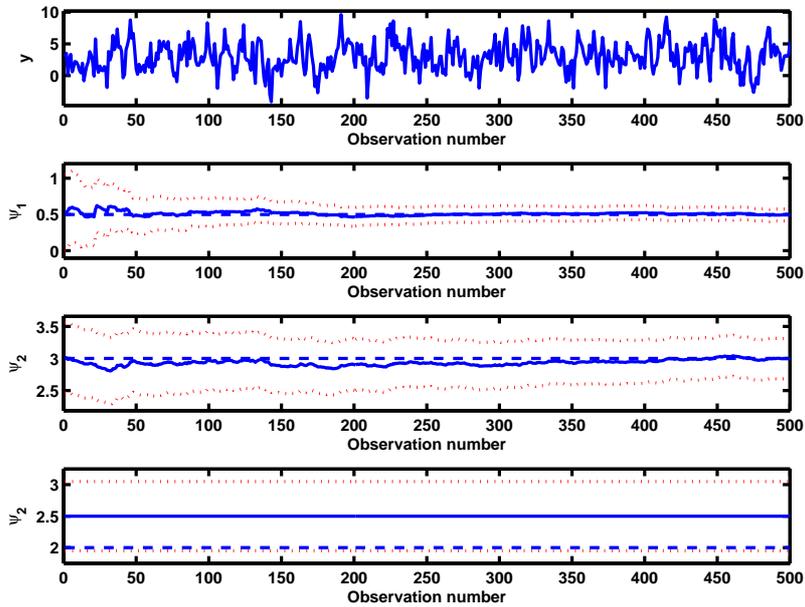}}
\caption{\label{abb.ukf_script1}Estimation of the parameter set
$[\tilde{y}_2, \tilde{y}_3, \tilde{y}_4]$ using the the UKF initialized with $[0.5,3,2.5]$. The true parameter values
are $[0.5,3,2]$ and denoted by the dashed lines. Confidence intervals are $\sqrt{3}$ times estimated standard deviations to
each side. The volatility parameter is not estimated. $y$ denotes the realization of the Ornstein-Uhlenbeck process.}
\end{center} \end{minipage} }
\end{center} \end{figure*}

Figures \ref{abb.ukf_script2} and \ref{abb.ukf_script1} show sequential estimates of these four state variables, where the true parameter values are  $\Psi=[\Psi_1,\Psi_2,\Psi_3]=[0.5,3,2]$.
The filters are initialized with the initial parameters
$\Psi_0=[0.5,3,2.5]$ and a diagonal initial covariance matrix with
standard deviations of $0.1,$ $0.1,$ $0.1$ for the parameters, respectively.
As expected, the UKF does not estimate the volatility parameter due to missing correlation between parameter and observations. As expected, the higher order UKF estimates the volatility parameter since volatility and squared prediction error are correlated. The estimation of the other parameters seems unaffected by the additional terms in the measurement update.
{\begin{figure*} [p!] \begin{center}
\makebox{ \begin{minipage}{0.65\linewidth} \begin{center}
\resizebox{\linewidth}{!}{\includegraphics*{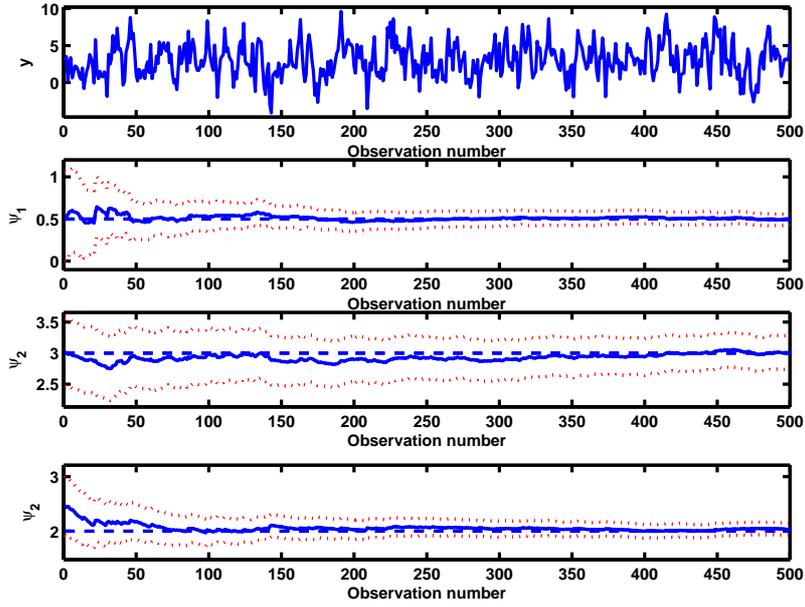}}
\caption{\label{abb.ukf_script2}Estimation of the parameter set
$[\Psi_1,\Psi_2,\Psi_3]$ using the higher order
correlation UKF initialized with $[0.5,3,2.5]$. The true parameter values
are $[0.5,3,2]$ and denoted by the dashed line. Confidence intervals are $\sqrt{3}$ times estimated standard deviations to
each side. $y$ denotes the realization of the Ornstein-Uhlenbeck process.}
\end{center} \end{minipage} }
\end{center} \end{figure*}}
A more detailed analysis of the filters is provided in table \ref{tab.HUKF_UKF}. Again, we sequentially estimate the model, but this time the true parameter values $\Psi=[\Psi_1,\Psi_2,\Psi_3]$ are drawn from a multivariate normal distribution,
  $$[\Psi_1,\Psi_2,\Psi_3] \sim \mathcal{N}\left([0.5,3,2], \text{diag}(0.1,0.1, 0.1)\right),$$
  i.e., the parameters are random but fixed. We simulate observations $z_i,$ $i=1,...,T$ with $T\in\{10,\dots,1000\}$ for $N=100000$ times, where each of the repetitions has a different parameter set. The observations are filtered with initial parameters (mean and covariance)
  $$\Psi_0   \sim \mathcal{N}\left([0.5,3,2], \text{diag}(0.1,0.1, 0.1)\right),$$
i.e., the true distribution of the parameters. After $T$ observations we get estimates (mean and variance) of each each parameter.

Table \ref{tab.HUKF_UKF} shows the filtering results obtained from HUKF and UKF. Four columns of results are shown for each filter. In the first column, the means of the deviations of the $N=100000$ estimates from the original parameter $m(\nu_{\Psi_j})=\text{mean}[\widehat{\Psi_j}-\Psi_j], j=1,2,3$ are shown. For unbiased estimates, this number should fluctuate around zero which is fulfilled in all cases. In the second column, the mean squared errors $m(\nu_{\Psi_j}^2)=\text{mean}[(\widehat{\Psi_j}-\Psi_j)^2], j=1,2,3$ are shown. The values  should be small. The values of the third column are the means of the estimated variances of the filters $m(\widehat{\sigma^2}_{\Psi_j})$ over the N=100000 simulations. Matching of the second and third column indicates that the estimates of the variances in the filter equations are unbiased.

The fourth column ($m_{0.95}$) presents the proportion of the cases where the absolute deviation of mean estimate $\widehat{\Psi_j}$ from the true value $\Psi_j$ is larger than $1.96\widehat{\sigma}_{\Psi_j}.$ For normally distributed estimators, these numbers should vary around $0.05.$ Since the distribution of in particular the volatility parameter $\Psi_3$ is not normal we expect deviations. However, in all cases the deviations are small.

Both filters show similar results for the parameters $\Psi_1$ and $\Psi_2.$ However, the results for the volatility parameter $\Psi_3$ deviate since the UKF does not update $\Psi_3$ and its variance after observations. Therefore, the third column (UKF) shows the initial value of the parameter variance ($0.1$) in all cases. The second column matches this value since the estimated values correspond to (always the same) initial parameter value $\Psi_{0;3}=2$ while the true parameter values $\Psi_3$ are drawn from a distribution with variance $0.1$ and mean $\Psi_{0;3}=2.$ The first column shows values around zero since the initial parameter $\Psi_{0;3}=2$ is an unbiased estimator of the true parameters (which are drawn from a normal distribution around $\Psi_{0;3}=2$). However, the UKF does not estimate this parameter. Contrary, the HUKF estimates this parameters very accurately and shows only a small bias for very small sample sizes which vanishes for sample sizes larger than $T=100.$ The estimated variances match the mean squared error and decrease rapidly with growing value of $T$.
{
\begin{table}[h!]
  \begin{center}
\begin{tabular}{cc|rrrr|rrrr}
& & & \multicolumn{2}{c}{HUKF} & & & \multicolumn{2}{c}{UKF} & \\ \hline \hline 
 & & & & & & & & & \\ $\psi_1=.5$ & T & $m(\nu_{\psi_1})$ & $m(\nu^2_{\psi_1})$ & $m(\widehat{\sigma^2}_{\psi_1})$ & $m_{0.95}$ & $m(\nu_{\psi_1})$ & $m(\nu^2_{\psi_1})$ & $m(\widehat{\sigma^2}_{\psi_1})$ & $m_{0.95}$ \\ \hline
&    10 & -.003 & .051 & .053 & .043 & -.003 & .051 & .053 & .044 \\ 
&    50 & .002 & .014 & .014 & .052 & .001 & .014 & .014 & .055 \\ 
&   100 & .001 & .007 & .007 & .051 & -.000 & .007 & .007 & .055 \\ 
&   250 & .000 & .003 & .003 & .050 & -.001 & .003 & .003 & .055 \\ 
&   500 & .000 & .001 & .001 & .051 & -.001 & .001 & .001 & .055 \\ 
&  1000 & .000 & .001 & .001 & .050 & -.001 & .001 & .001 & .055 \\ 
 & & & & & & & & & \\ $\psi_2=3$ & T & $m(\nu_{\psi_2})$ & $m(\nu^2_{\psi_2})$ & $m(\widehat{\sigma^2}_{\psi_2})$ & $m_{0.95}$ & $m(\nu_{\psi_2})$ & $m(\nu^2_{\psi_2})$ & $m(\widehat{\sigma^2}_{\psi_2})$ & $m_{0.95}$ \\ \hline
&    10 & -.001 & .096 & .094 & .052 & -.001 & .096 & .094 & .052 \\ 
&    50 & -.000 & .079 & .074 & .056 & -.000 & .079 & .075 & .056 \\ 
&   100 & -.001 & .066 & .061 & .058 & -.001 & .067 & .062 & .059 \\ 
&   250 & -.001 & .048 & .043 & .060 & -.001 & .048 & .044 & .059 \\ 
&   500 & .000 & .036 & .032 & .060 & .000 & .036 & .033 & .061 \\ 
&  1000 & .000 & .026 & .023 & .057 & .000 & .026 & .023 & .060 \\ 
 & & & & & & & & & \\ $\psi_3=2$ & T & $m(\nu_{\psi_3})$ & $m(\nu^2_{\psi_3})$ & $m(\widehat{\sigma^2}_{\psi_3})$ & $m_{0.95}$ & $m(\nu_{\psi_3})$ & $m(\nu^2_{\psi_3})$ & $m(\widehat{\sigma^2}_{\psi_3})$ & $m_{0.95}$ \\ \hline
&    10 & -.034 & .073 & .071 & .055 & -.000 & .099 & .100 & .049 \\ 
&    50 & -.016 & .030 & .031 & .048 & -.001 & .100 & .100 & .050 \\ 
&   100 & -.009 & .017 & .018 & .048 & .001 & .099 & .100 & .049 \\ 
&   250 & -.004 & .008 & .008 & .048 & .001 & .099 & .100 & .048 \\ 
&   500 & -.002 & .004 & .004 & .047 & -.000 & .100 & .100 & .050 \\ 
&  1000 & -.001 & .002 & .002 & .049 & -.001 & .100 & .100 & .050 \\ 
 \end{tabular}

 \end{center}
 \caption{\label{tab.HUKF_UKF} Sequential estimation of the three parameters $\Psi_j, j=1,2,3$ of model (\ref{glg.zustmodellsimu2}) with the higher order filter (HUKF) and the unscented Kalman filter (UKF) for different length $T$ of time series. The estimates are repeated $N=100000$ times with true parameter values drawn from a multivariate normal distribution. Shown are the mean errors of estimation ($m(\nu_{\Psi_j})$), mean squared errors of the estimates ($m(\nu_{\Psi_j}^2)$), the mean of the estimated parameter variances ($m(\widehat{\sigma^2}_{\Psi_j})$) and the empirical exceedance ratios ($m_{0.95}$) of the estimated confidence intervals of the parameters (1.96 standard deviations to both sides).
}
\end{table}
}

\section{Conclusion}
\label{sec.concl}  This paper addresses sequential estimation of parameters and states in nonlinear state-space models in which the parameters are not correlated with the observations, as, for example, volatility. It is focused on nonlinear, Gaussian stochastic filters, i.e., filters that reduce all densities to their first two moments. These filters, like the extended Kalman filter or the unscented Kalman filter, are fast filters and give good results in many cases. However, when there is no correlation between states and observations, they fail to estimate the respective states and parameters. In this paper, higher order correlation filter equations are introduced, that are also based on Gaussian densities but do not rely on linear correlations. The approach shows large advantages over simulation-based filtering
methods with respect
to computing costs. Filter algorithms for a higher order unscented Kalman filter are explicitly stated. The filter is validated in a simulation study.

\newpage
\bibliographystyle{elsarticle-num}


\begin{thebibliography}{10}
\expandafter\ifx\csname url\endcsname\relax
  \def\url#1{\texttt{#1}}\fi
\expandafter\ifx\csname urlprefix\endcsname\relax\def\urlprefix{URL }\fi
\expandafter\ifx\csname href\endcsname\relax
  \def\href#1#2{#2} \def\path#1{#1}\fi

\bibitem{Kalman1960}
R.~E. Kalman, A new approach to linear filtering and prediction problems,
  Transactions of the ASME--Journal of Basic Engineering 82~(Series D) (1960)
  35--45.

\bibitem{Tanizaki1996}
H.~Tanizaki, Nonlinear Filters. Estimation and Applications, Springer-Verlag,
  1996.

\bibitem{HermosoCarazo}
A.~Hermoso-Carazo, J.~Linares-Pérez, Different approaches for state filtering
  in nonlinear systems with uncertain observations, Applied Mathematics and
  Computation 187~(2) (2007) 708 -- 724.

\bibitem{Jazwinski1970}
A.~H. Jazwinski, Stochastic Processes and Filtering Theory, Academic Press, New
  York, 1970.

\bibitem{Julier1997}
S.~Julier, J.~K. Uhlmann, A new extension of the kalman filter to nonlinear
  systems, Proc. of AeroSense: The 11th Int. Symp. A.D.S.S.C.

\bibitem{Pitt1999}
M.~K. Pitt, N.~Shephard, Filtering via simulation: Auxiliary particle filters,
  Journal of The American Statistical Association 94~(446) (1999) 590--599.

\bibitem{Singer2003}
H.~Singer, Nonlinear continuous-discrete filtering using kernel density
  estimates and functional integrals, Journal of Mathematical Sociology 27
  (2003) 1--28.

\bibitem{Julier2004}
S.~Julier, J.~K. Uhlmann, Unscented filtering and nonlinear estimation, Proc.
  of the IEEE 92~(3).

\bibitem{Godambe1991}
V.~P. Godambe (Ed.), Estimating Functions, Oxford University Press, 1991.

\bibitem{NaikNimbalkar1995}
U.~V. Naik-Nimbalkar, M.~B. Rajarshi, Filtering and smoothing via estimating
  functions, Journal of The American Statistical Association 90~(429) (1995)
  301--306.

\bibitem{Thompson1999}
M.~E. Thompson, A.~Thavaneswaran, Filtering via estimating functions, Applied
  Mathematics Letters 12~(5) (1999) 61--67.

\bibitem{Crowder1987}
M.~Crowder, On linear and quadratic estimating functions, Biometrika 74~(3)
  (1987) 591--597.

\bibitem{Crowder1986}
M.~Crowder, On consistency and inconsistency of estimating equations,
  Econometric Theory 2~(3) (1986) 305--330.

\bibitem{Godambe1989137}
V.~P. Godambe, M.~E. Thompson, An extension of quasi-likelihood estimation,
  Journal of Statistical Planning and Inference 22~(2) (1989) 137--152.

\bibitem{Heyde1987}
C.~Heyde, On combining quasi-likelihood estimating functions, Stochastic
  Processes and their Applications 25 (1987) 281--287.

\bibitem{Wefelmeyer1996}
W.~Wefelmeyer, Quasi-likelihood models and optimal inference, The Annals of
  Statistics 24~(1) (1996) 405--422.

\bibitem{McNamee1967}
J.~McNamee, F.~Stenger, Construction of fully symmetric numerical integration
  formulas, Numerische Mathematik 10 (1967) 327--344.

\bibitem{Lerner2002}
U.~N. Lerner, Hybrid Bayesian Networks for Reasoning About Complex Systems,
  Ph.D. thesis, Stanford Univ., 2002.

\bibitem{Schwartz1997}
E.~S. Schwartz, The stochastic behavior of commodity prices: Implications of
  valuation and hedging, The Journal of Finance 52~(3) (1997) 923--973.

\bibitem{Lucia2002}
J.~J. Lucía, E.~S. Schwartz, Electricity process and power derivates: Evidence
  from the nordic power exchange, Review of Derivates Research 5 (2002) 5--50.

\bibitem{Cox1985}
J.~C. Cox, J.~E. Ingersoll, S.~A. Ross, A theory of the term structure of
  interest rates, Econometrica 53 (1985) 385--407.

\end{thebibliography}

\end{document}